# Electric charge quantization in $SU(3)_C \times SU(3)_L \times U(1)_X$ model

**O. B. Abdinov, F.T. Khalil-zade, S. S. Rzaeva**
Institute of Physics of Azerbaijan National Academy of Sciences, AZ143, Baku, G. Javid av.33

E – mail: abdinov@physics.ab.az
fkhalilzade@physics.ab.az
sevda_hep@physics.ab.az

**Abstract.** Basing on the general photon eigenstate and anomaly cancellation, it is shown that the electric charge quantization in $SU(3)_C \times SU(3)_L \times U(1)_X$ model with exotic particles can be obtained independently on parameters α and β. The fixation of hypercharges of fermions fields by the Higgs fields and dependence of the electric charges quantization conditions from the hypercharges of Higgs fields leads to the fact that the electric charge in the considered model can be quantized and fixed only in the presence of Higgs fields. In addition, we have shown that in the considered model the classical constraints following from the Yukawa interactions are equivalent to the conditions following from the P – invariance of electromagnetic interaction. The most general expressions for the gauge bosons masses, eigenstates of neutral fields and the interactions of leptons and quarks with gauge bosons have been derived in the arbitrary case.

## I. Introduction

Standard Model (SM) of strong and electroweak interaction well describing existing experimental data involves several unanswered questions. Within SM have not been solved such problems as existence of three families [1, 2], mass hierarchy problem [1, 2], electric charge quantization etc. Solution of mentioned theoretical problems can be achieved either by introduction of additional particles or by enlargement of symmetry group. For instance, the SU(5) grand unification model [3] can unify the interactions and predicts the electric charge quantization, while the models based on $E_6$ group can also unify the interactions and might explain the masses of the neutrinos [4].

Interesting alternative to explain the origin of generations comes from the cancellation of chiral anomalies [5]. In particular, the models with gauge group $SU(3)_C \times SU(3)_L \times U(1)_X$ [6-9], (called 3-3-1 models) arise as a possible solution to this puzzle, since some of such models require the three generations in order to cancel chiral anomalies completely [9-11]. An additional motivation to study this kind of models comes from the fact that in these models there are some progress in investigations of such problems as neutrino mass [12], P – parity violation in nuclear transitions [13] and etc. Electric charge quantization in two model based on the $SU(3)_C \times SU(3)_L \times U(1)_X$ gauge group, namely in the minimal model and in the model with right-handed neutrino has been considered in [14]. Authors have shown, that electric charge quantization is not dependent on the classical constraints on generating mass to the fermions, is related closely with the generation number problem and is a direct consequence of the fermion content under the anomaly free conditions.

It should be noted that the electric charge quantization problem considered in [14] has been derived for the concrete values of parameters α and β. Besides in these type models based on $SU(3)_C \times SU(3)_L \times U(1)_X$ group symmetry [5-14], the expressions for the masses of neutral gauge bosons and eigenstates of neutral fields have been obtained in particular case.

It has been shown [15, 16] that photon eigenstate depends from the hypercharges of Higgs fields (see, also [14]), that leads to necessity of more detailed research of electric charge quantization in gauge theories. This work is devoted to investigation of electric charge quantization in $SU(3)_C \times SU(3)_L \times U(1)_X$ model with exotic particles independent of parameters α and β.

## II. Model structure

The electric charge is defined in general as a linear combination of the diagonal generators of $SU(3)_C \times SU(3)_L \times U(1)_X$ group

$$\hat{Q} = \alpha \hat{T}_3 + \beta \hat{T}_8 + X \hat{I}, \qquad (1)$$

with $T_3 = diag(1,-1,0)/2$ and $T_8 = diag(1,1,-2)/2\sqrt{3}$, where the normalization chosen is $Tr(T_\alpha T_\beta) = \delta_{\alpha\beta}/2$ and $I = diag(1,1,1)$ is the identity matrix. The value of the parameters α and β determines the fermion assignment and it is customary to use this number to classify the different models (see, for example [17]).

The hypercharges of fermions (as well as the Higgs) fields causing interaction with Maxwell field, is defined as

$$\hat{Y} = \beta \hat{T}_8 + X \hat{I}. \qquad (2)$$

Note that as the aim of this paper is the study of electric charge quantization, the expressions of electric charge (1) and hypercharge (2) further are not used.

Let's consider the case when symmetry is broken by the Higgs fields

$$<\chi> = \frac{1}{\sqrt{2}} \begin{pmatrix} 0 \\ 0 \\ V \end{pmatrix}, \quad <\rho> = \frac{1}{\sqrt{2}} \begin{pmatrix} 0 \\ v \\ 0 \end{pmatrix}, \quad <\eta> = \frac{1}{\sqrt{2}} \begin{pmatrix} u \\ 0 \\ 0 \end{pmatrix}. \qquad (3)$$

The part of interaction Lagrangian responsible for the Higgs fields looks like

$$V_{kin} = (D_\mu \chi)^+ (D_\mu \chi) + (D_\mu \eta)^+ (D_\mu \eta) + (D_\mu \rho)^+ (D_\mu \rho). \qquad (4)$$

The covariant derivative $D_\mu$ is given by

$$D_\mu = \partial_\mu - igT_a W_{a\mu} - ig'T_9 X B_\mu, \qquad (5)$$

where $T_a$ $(a = 1,...,8)$ are the $SU(3)_L$ generators, and $T_9 = diag(1,1,1)/\sqrt{6}$ are defined as $Tr(T_a T_b) = \delta_{ab}/2$, $(a,b = 1,2,...,9)$; $g$ and $g'$ – coupling constants.

To keep consistency with the effective theory, the VEVs in the model satisfy the constraint: $V \gg v \gg u$ (similarly to papers [7]).

For lepton and quark fields we choose the flowing representations (we will consider one family of leptons and quarks without mixing):

$$\psi_{lL} = \begin{pmatrix} v_e \\ e^- \\ N \end{pmatrix}_L \sim (1,3,y_{lL}), \quad \psi_{eR} = e_R \sim (1,1,y_{eR}), \quad \psi_{NR} = N_R \sim (1,1,y_{NR}),$$

$$\psi_{QL} = \begin{pmatrix} u \\ d \\ U \end{pmatrix}_L \sim (3,3,y_{QL}), \quad \psi_{uR} = u_R \sim (3,1,y_{uR}), \quad \psi_{dR} = d_R \sim (3,1,y_{dR}), \quad \psi_{UR} = U_R \sim (3,1,y_{UR}). \qquad (6)$$

### III. Masses of gauge bosons

The gauge bosons of this model form an octet $W_{a\mu}$ and singlet $B_\mu$ associated with $SU(3)_L$ and $U(1)$ accordingly. It is easy to see that the massless gauge bosons associated with $SU(3)_C$ group decouple from the neutral bosons mass matrix and are not considered late. The masses matrix of gauge bosons arises from the kinetic part (4) of interaction lagrangian. The covariant derivatives for the triplet of Higgs fields look likes

$$D_\mu \varphi_i = \partial_\mu \varphi_i - i P_\mu \varphi_i, \qquad (7)$$

were $\varphi_i$ - Higgs fields (3), and matrix $P_\mu$ has the form

$$P_\mu = \frac{g}{2} \begin{pmatrix} W_{3\mu} + \frac{W_{8\mu}}{\sqrt{3}} + \sqrt{\frac{2}{3}} t X B_\mu & \sqrt{2} W_\mu^+ & \sqrt{2} X_\mu^{\prime 0} \\ \sqrt{2} W_\mu^- & -W_{3\mu} + \frac{W_{8\mu}}{\sqrt{3}} + \sqrt{\frac{2}{3}} t X B_\mu & \sqrt{2} Y_\mu^{\prime -} \\ \sqrt{2} X_\mu^{\prime 0 *} & \sqrt{2} Y_\mu^{\prime +} & -\frac{2 W_{8\mu}}{\sqrt{3}} + \sqrt{\frac{2}{3}} t X B_\mu \end{pmatrix}, \qquad (8)$$

where $t = g'/g$ and

$$W_\mu^\pm = \frac{W_{1\mu} \mp i W_{2\mu}}{\sqrt{2}}, \quad Y_\mu^{\prime \mp} = \frac{W_{6\mu} \mp i W_{7\mu}}{\sqrt{2}}, \quad X_\mu^{\prime 0} = \frac{W_{4\mu} - i W_{5\mu}}{\sqrt{2}}. \qquad (9)$$

In this case taking into account (3), (8) and (9) in (4) for the masses of gauge bosons we have

$$L_{mass} = M_X^2 X_\mu^{\prime 0} X_\mu^{\prime 0*} + M_W^2 W_\mu^+ W_\mu^- + M_Y^2 Y_\mu^{\prime +} Y_\mu^{\prime -} + \frac{g^2 u^2}{8}\left(W_{3\mu} + \frac{1}{\sqrt{3}} W_{8\mu} + \sqrt{\frac{2}{3}} t X_\eta B_\mu\right)^2 +$$

$$+ \frac{g^2 v^2}{8}\left(-W_{3\mu} + \frac{1}{\sqrt{3}} W_{8\mu} + \sqrt{\frac{2}{3}} t X_\rho B_\mu\right)^2 + \frac{g^2 V^2}{8}\left(-\frac{2}{\sqrt{3}} W_{8\mu} + \sqrt{\frac{2}{3}} t X_\chi B_\mu\right)^2, \qquad (10)$$

where

$$M_W^2 = \frac{g^2}{4}\left(v^2 + u^2\right), \quad M_Y^2 = \frac{g^2}{4}\left(V^2 + v^2\right), \quad M_X^2 = \frac{g^2}{4}\left(V^2 + u^2\right). \qquad (11)$$

Note that as far as the masses (11) and interactions of gauge bosons (9) are not the subject of this paper we will not discuss them later. Discussion of these problems can be found in [5-14].

The interactions lagrangian, containing the mass of the neutral gauge bosons in this case, looks like:

$$L_{mass}^{NG} = \frac{1}{2} V^T M_0^2 V, \qquad (12)$$

where $V^T = (W_{3\mu}, W_{8\mu}, B_\mu)$ and

$$M_0^2 = \frac{g^2}{4}\begin{pmatrix} m_{11} & m_{12} & m_{13} \\ m_{12} & m_{22} & m_{23} \\ m_{13} & m_{23} & m_{33} \end{pmatrix}, \qquad (13)$$

here

$$m_{11} = (u^2 + v^2), \quad m_{12} = \frac{1}{\sqrt{3}}(u^2 - v^2), \quad m_{13} = \frac{2}{\sqrt{6}}t(u^2 X_\eta - v^2 X_\rho), \quad m_{22} = \frac{1}{3}(4V^2 + u^2 + v^2),$$

$$m_{23} = \frac{2}{3\sqrt{2}}t(u^2 X_\eta + v^2 X_\rho - 2V^2 X_\chi), \quad m_{33} = \frac{2}{3}t^2(u^2 X_\eta^2 + v^2 X_\rho^2 + V^2 X_\chi^2). \tag{14}$$

The eigenvalues of the mass matrix (13) are the roots of equation

$$M^3 - \chi_1 M^2 + \chi_2 M - \chi_3 = 0, \tag{15}$$

where $M = 4M_0^2/g^2$ and

$$\chi_1 = \frac{2}{3}\left[2(u^2 + v^2 + V^2) + t^2(u^2 X_\eta^2 + v^2 X_\rho^2 + V^2 X_\chi^2)\right],$$

$$\chi_2 = \frac{4}{3}\left\{(u^2 V^2 + v^2 V^2 + u^2 v^2) + \frac{2t^2}{3}\left[u^2 V^2(X_\eta^2 + X_\chi^2 + X_\eta X_\chi) + v^2 V^2(X_\rho^2 + X_\chi^2 + X_\rho X_\chi) + u^2 v^2(X_\eta^2 + X_\rho^2 + X_\eta X_\rho)\right]\right\},$$

$$\chi_3 = \frac{8}{9}u^2 v^2 V^2 t^2 (X_\eta + X_\rho + X_\chi)^2. \tag{16}$$

In general case the eigenvalues of mass matrix corresponding to the masses of neutral gauge bosons can be real and differ from zero. In the case when the roots of equation (15) satisfy condition $M_1^2 \gg M_2^2 \gg M_3^2$ (that is in agreement with experimental data [19]), we have

$$M_{Z_1}^2 \approx \frac{g^2 V^2}{6}(2 + t^2 X_\chi), \quad M_{Z_2}^2 \approx \frac{g^2 v^2}{6} \cdot \frac{3 + 2t^2(X_\rho^2 + X_\chi^2 + X_\rho X_\chi)}{2 + t^2 X_\chi},$$

$$M_{Z_3}^2 \approx \frac{g'^2 u^2}{2} \frac{(X_\eta + X_\rho + X_\chi)^2}{3 + 2t^2(X_\rho^2 + X_\chi^2 + X_\rho X_\chi)}. \tag{17}$$

In our case ($V \gg v \gg u$) the easiest mass in (17) can be identified with photon mass ($M_3^2 = M_\gamma^2$) only under the condition

$$X_\eta + X_\rho + X_\chi = 0. \tag{18}$$

Notice that this condition also follows from the electric charge conservation [14]. In this case for the masses of neutral bosons we have

$$M_\gamma^2 = 0, \quad M_{Z_{1,2}}^2 \approx \frac{g^2}{8}\left[\chi_1 \pm \sqrt{\chi_1^2 - 4\chi_2}\right]. \tag{19}$$

The easiest of massive neutral vector bosons (19), can be identified whit SM $Z$ - boson $M_{Z_2} \equiv M_Z$.

### IV. Electric charge quantization

Transformation of neutral fields $W_{3\mu}, W_{8\mu}, B_\mu$ to the physical photon field, can be written in the form

$$A_\mu = a_1 W_{3\mu} + a_2 W_{8\mu} + a_3 B_\mu. \tag{20}$$

The eigenstate with zero eigenvalue follow from the equation

$$M_0^2 \begin{pmatrix} a_1 \\ a_2 \\ a_3 \end{pmatrix} = 0. \tag{21}$$

It can be checked that (taking into account (18)) the matrix $M_0^2$ has a non-degenerate zero eigenvalue and corresponding eigenstate can be identified with physical photon field $A_\mu$.

In the considering model for the quantities $a_i$ $(i = 1 \div 3)$ we have

$$a_1 = -\frac{g'}{\sqrt{2}\bar{g}}(X_\rho - X_\eta), \quad a_2 = \frac{3g'}{\sqrt{6}\bar{g}}(X_\rho + X_\eta) \quad a_3 = -\frac{\sqrt{3}g}{\bar{g}}, \tag{22}$$

where

$$\bar{g} = g[3 + 2t^2(X_\eta^2 + X_\rho^2 + X_\eta X_\rho)]^{1/2}. \tag{23}$$

From the expressions (20), (22) and (23) one can see that, the photon eigenstate is independent on VEVs structure. This is a natural consequence of the $U(1)$ invariance [14-16]. However photon eigenstate depends from the Higgs fields hypercharges. Moreover, to be consistent with the QED based on the unbroken $U(1)$ gauge group, the photon field has to keep the general properties of the electromagnetic interaction in the framework of the 3-3-1 model, such as the parity invariant nature [20]. These would help us to obtain some consequences related to quantities which are independent on VEVs structure.

At first let us consider interaction of leptons with the electromagnetic field. In the considered model it looks like

$$L_{l\gamma} = Q_\nu \bar{\nu}_e \gamma_\mu (1 + \gamma_5) \nu_e A_\mu + \bar{e} \gamma_\mu (Q_{0e} + Q'_{0e} \gamma_5) e A_\mu + \bar{N} \gamma_\mu (Q_N + Q'_N \gamma_5) N A_\mu, \tag{24}$$

where

$$Q_\nu = \frac{g}{4}[a_1 + \frac{1}{\sqrt{3}}a_2 + \sqrt{\frac{2}{3}}ta_3 y_{lL}], \quad Q_{0e} = \frac{g}{4}[-a_1 + \frac{1}{\sqrt{3}}a_2 + \sqrt{\frac{2}{3}}ta_3(y_{lL} + y_{eR})],$$

$$Q'_{0e} = \frac{g}{4}[a_1 + \frac{1}{\sqrt{3}}a_2 + \sqrt{\frac{2}{3}}ta_3(y_{lL} - y_{eR})], \quad Q_N = \frac{g}{4}[-\frac{2}{\sqrt{3}}a_2 + \sqrt{\frac{2}{3}}ta_3(y_{lL} + y_{NR})],$$

$$Q'_N = -\frac{g}{4}[-\frac{2}{\sqrt{3}}a_2 + \sqrt{\frac{2}{3}}ta_3(y_{lL} - y_{NR})]. \tag{25}$$

Taking into account the parity invariance of the electromagnetic interaction from (24), we have

$$Q_\nu = 0, \quad Q'_{0e} = 0, \quad Q'_N = 0. \tag{26}$$

In the considered case when neutrino has not the right component, the requirement parity invariance of electromagnetic interaction and the condition of neutrino charge equality to zero are equivalent. Besides, from the condition of parity invariance of electromagnetic interaction we have the relations between hypercharges of Higgs and lepton fields

$$y_{lL} = X_\eta, \quad y_{eR} = X_\eta - X_\rho, \quad y_{NR} = X_\eta - X_\chi. \tag{27}$$

Consequently for the electric charges of leptons we have

$$Q_\nu = 0, \quad Q_{0e} = -Q_e, \quad Q_N = -Q_e \frac{2X_\eta + X_\rho}{X_\eta - X_\rho}, \tag{28}$$

where

$$Q_e = \frac{gg'}{\sqrt{2}\bar{g}}(X_\eta - X_\rho). \tag{30}$$

In the considered model the Yukawa interactions which induce masses for the leptons can be written as

$$L_Y^l = f_e \bar{\psi}_{lL} \rho \psi_{eR} + f_N \bar{\psi}_{lL} \chi \psi_{NR} + h.c. \tag{31}$$

From (31) under the U (1) invariance we also have conditions (27). As a result we can conclude that conditions following from the P-invariance of electromagnetic interaction are equivalent to the conditions following from the Yukawa interactions which induce masses for the leptons. The equations (27) are the fixing condition for the hypercharges of the leptons fields by the Higgs fields and further it will be shown that they are also the conditions of electric charge quantization of leptons.

Let's consider interaction of quarks with electromagnetic field. In the considered model it looks like

$$L_{q\gamma} = \bar{u}\gamma_\mu(Q_u + Q'_u\gamma_5)uA_\mu + \bar{d}\gamma_\mu(Q_d + Q'_d\gamma_5)dA_\mu + \bar{U}\gamma_\mu(Q_U + Q'_U\gamma_5)UA_\mu, \qquad (32)$$

where

$$Q_u = \frac{g}{4}[a_1 + \frac{1}{\sqrt{3}}a_2 + \sqrt{\frac{2}{3}}ta_3(y_{QL} + y_{uR})], \quad Q'_u = \frac{g}{4}[a_1 + \frac{1}{\sqrt{3}}a_2 + \sqrt{\frac{2}{3}}ta_3(y_{QL} - y_{uR})],$$

$$Q_d = \frac{g}{4}[-a_1 + \frac{1}{\sqrt{3}}a_2 + \sqrt{\frac{2}{3}}ta_3(y_{QL} + y_{dR})], \quad Q'_d = \frac{g}{4}[-a_1 + \frac{1}{\sqrt{3}}a_2 + \sqrt{\frac{2}{3}}ta_3(y_{QL} - y_{dR})],$$

$$Q_U = \frac{g}{4}[-\frac{2}{\sqrt{3}}a_2 + \sqrt{\frac{2}{3}}ta_3(y_{QL} + y_{UR})], \quad Q'_U = \frac{g}{4}[-\frac{2}{\sqrt{3}}a_2 + \sqrt{\frac{2}{3}}ta_3(y_{QL} - y_{UR})]. \qquad (33)$$

Similarly to the of leptons case taking into account P-invariance of electromagnetic interaction from (32) and (33) we have conditions

$$Q'_u = 0, \quad Q'_d = 0, \quad Q'_U = 0. \qquad (34)$$

These conditions lead to the following relations between hypercharges of Higgs and quarks fields

$$y_{QL} - y_{uR} = X_\eta, \qquad y_{QL} - y_{dR} = X_\rho, \qquad y_{QL} - y_{UR} = X_\chi. \qquad (35)$$

The expressions (35) fix the left and right hypercharges difference of quarks fields are also the conditions of quarks electric charge quantization. Equations (35) also follow from the Yukawa interactions which induce masses for the quarks

$$L_Y^q = f_u\bar{\psi}_{QL}\eta\psi_{uR} + f_d\bar{\psi}_{QL}\rho\psi_{dR} + f_U\bar{\psi}_{QL}\chi\psi_{UR} + h.c, \qquad (36)$$

Similarly to the leptons case we can conclude that conditions following from the P-invariance of electromagnetic interaction are equivalent to the conditions following from the Yukawa interactions which induce masses for the quarks. (See also [15, 16]). Taking into account (35), (22) and (18) in (33) for the quarks electric charges we have

$$Q_u = Q_e \frac{X_\eta - y_{QL}}{X_\eta - X_\rho} \qquad Q_d = Q_e \frac{X_\rho - y_{QL}}{X_\eta - X_\rho} \qquad Q_U = Q_e \frac{X_\chi - y_{QL}}{X_\eta - X_\rho} \qquad (37)$$

The obtained expressions (28) and (37) can be considered as the evidence of electric charge quantization of leptons and quarks. However these expressions do not define numerical values of electric charges of leptons and quarks (in terms of electron charge). For obtaining of the numerical values for the leptons and quarks electric charges, it is necessary to have the additional relations between fermions field hypercharges. Such of relations can be obtained from the conditions of cancellations of gauge [5, 21] and mixed gauge-gravitational anomalies [22]. In the considered model we have

$$\begin{aligned} y_{lL} + 3y_{QL} &= 0, \\ 3y_{QL} - y_{uR} - y_{dR} - y_{UR} &= 0, \\ 3y_{lL} + 9y_{QL} - 3(y_{uR} + y_{dR} + y_{UR}) - y_{eR} - y_{NR} &= 0, \\ 3y_{lL}^3 + 9y_{QL}^3 - 3(y_{uR}^3 + y_{dR}^3 + y_{UR}^3) - y_{eR}^3 - y_{NR}^3 &= 0. \end{aligned} \qquad (38)$$

From the first equation (38) (with taken into account (27)) we have:

$$y_{QL} = -\frac{1}{3}X_\eta, \qquad y_{uR} = -\frac{4}{3}X_\eta \qquad y_{dR} = -\frac{1}{3}X_\eta - X_\rho, \qquad y_{UR} = -\frac{1}{3}X_\eta - X_\chi. \qquad (39)$$

Taking into account (35) from the second and third equations (38) we have expression (18). Fourth equation (38) leads to:

$$X_\eta = -X_\rho. \tag{40}$$

Consequently for the hypercharges of fermions fields we have:

$$y_{lL} = X_\eta, \qquad y_{eR} = 2X_\eta \qquad y_{NR} = X_\eta.$$

$$y_{QL} = -\frac{1}{3}X_\eta, \qquad y_{uR} = -\frac{4}{3}X_\eta, \qquad y_{dR} = \frac{2}{3}X_\eta, \qquad y_{UR} = -\frac{1}{3}X_\eta. \tag{41}$$

This leads to the electric charge quantization

$$Q_\nu = 0, \qquad Q_e = \frac{\sqrt{2}gg'X_\eta}{(3+2X_\eta^2 t^2)^{1/2}}, Q_N = -\frac{1}{2}Q_e, Q_u = \frac{2}{3}Q_e, Q_d = -\frac{1}{3}Q_e, Q_U = \frac{1}{6}Q_e, \tag{42}$$

Similar expressions can be written for other fermions. Conditions (27) and (35) fix the hypercharges of fermions fields. The conditions following from the anomalies cancellations (taking into account (27) and (35)), fix hypercharges of all remained fields. Thus, if there are no conditions (27) and (35) it is obvious that to solve the equations following from the anomalies cancellations is impossible and consequently there are not electric charge quantizations, hence, these conditions are electric charge quantization ones. However these conditions depend from the hypercharges of Higgs fields, so these facts can be interpreted as a presence of influence of Higgs fields on the electric charge quantization.

It is necessary to note that unlike results of work [14] in which authors have shown that electric charge quantization dos not depend on the classical constraints on generating mass to the fermions in the considered case the conditions following from the Yukawa interactions are equivalent to the conditions following from the P – invariance of electromagnetic interaction. Equivalence of conditions following from the P – invariance of electromagnetic interaction and from the classical constraints on generating mass to the fermions in the SM and $SU(3)_C \times SU(3)_L \times U(1)_X \times U'(1)_{X'}$ model has been shown in [15,16]. This fact may be useful for the explanation of the P – invariance of electromagnetic interaction.

### VI. Charged and neutral currents

For the eigenstate with nonzero eigenvalue we have

$$Z_{1\mu} = b_1 W_{3\mu} + b_2 W_{8\mu} + b_3 B_\mu,$$
$$Z_{2\mu} = c_1 W_{3\mu} + c_2 W_{8\mu} + c_3 B_\mu, \tag{43}$$

where

$$b_1 = \frac{\sqrt{2}g'\Delta_1}{\bar{g}_Z}, b_2 = \frac{\sqrt{6}g'\Delta_2}{\bar{g}_Z}, b_3 = -\frac{\sqrt{3}g\Delta_3}{\bar{g}_Z}. \tag{44}$$

Here (in the case of arbitrary Higgs field hypercharges)

$$\Delta_1 = (u^2 X_\eta - v^2 X_\rho)(4V^2 - \frac{12 M_{Z_1}^2}{g^2}) + 2u^2 v^2 (X_\eta - X_\rho) + 2V^2(u^2 - v^2)X_\chi,$$

$$\Delta_2 = 2u^2 v^2 (X_\eta + X_\rho) - 2V^2(u^2 + v^2)X_\chi - \frac{4M_{Z_1}^2}{g^2}(v^2 X_\rho + u^2 X_\eta - 2V^2 X_\chi), \tag{45}$$

$$\Delta_3 = -4V^2(u^2 + v^2) + \frac{16 M_{Z_1}^2}{g^2}(u^2 + v^2 + V^2) - \frac{48 M_{Z_1}^2}{g^4}, \quad \bar{g}_Z = \left[3g^2 \Delta_3^2 + 2g'^2(\Delta_1^2 + 3\Delta_2^2)\right]^{1/2}.$$

The magnitudes $c_i$ $(i = 1 \div 3)$ can be found from the appropriate expressions $b_i$ by substitution $Z_1 \to Z_2$.

In the most general form the interaction lagrangian of fermions with gauge bosons has the following form:

$$L_{int} = i\overline{\psi}_{fL}\gamma_\mu(\partial_\mu - ig\sum_{a=1}^{8}T_a W_{a\mu} - ig'T_9 XB_\mu)\psi_{fL} + i\overline{\psi}_{fR}\gamma_\mu(\partial_\mu - ig'XB_\mu)\psi_{fR}, \quad (46)$$

where $\psi_{fL}$, $\psi_{fR}$ – are left and right fermions fields.

Taking into account (43) and (46) we have

$$L_{int} = L_f^{CC} + L_f^{NC}, \quad (47)$$

where

$$L_f^{CC} = \frac{g}{\sqrt{2}}(\overline{\nu}_e W_\mu e_L + \overline{N}_L Y'_\mu e_L + \overline{\nu}_N X'_\mu N_L + \overline{d}_L W_\mu u_L + \overline{U}_L Y'_\mu d_L + \overline{u} X'_\mu U_L + h.c.). \quad (48)$$

$$L_f^{NC} = \frac{g}{4}\sum_f \overline{f}\gamma_\mu(g_{V_1}^f + g_{A_1}^f\gamma_5)f Z_{1\mu} + \frac{g}{4}\sum_f \overline{f}\gamma_\mu(g_{V_2}^f + g_{A_2}^f\gamma_5)f Z_{2\mu}, \quad (49)$$

where $f$ takes values $\nu_e, e, N, d, u, U$. For the coupling constants we have:

$$g_{V_1,A_1}^l = k_1^l b_1 + \frac{k_2^l}{\sqrt{3}}b_2 + \sqrt{\frac{2}{3}}tb_3(y_{lL} \pm y_{lR}), \quad g_{V_1,A_1}^q = k_1^q b_1 + \frac{k_2^q}{\sqrt{3}}b_2 + \sqrt{\frac{2}{3}}tb_3(y_{QL} \pm y_{qR}), \quad (50)$$

where upper signs relate to victories coupling constants and lower signs relates to axial ones. Besides:

for $l = \nu$    $k_1^\nu = k_2^\nu = 1$;    $y_{\nu R} = 0$; for $l = e$    $k_1^e = -1$,    $k_2^e = 1$; for $l = N$    $k_1^N = 0$,    $k_2^N = -1$;

for $q = u$    $k_1^u = k_2^u = 1$; for $q = d$    $k_1^d = -1$,    $k_2^d = 1$; for $q = U$    $k_1^U = 0$,    $k_2^U = 1$.

Note that the expression of magnitudes $g_{V_2}^f$ и $g_{A_2}^f$ – can be obtained from appropriate expressions (50) by the substitution $b_i \to c_i$.

### VII Conclusions

As a result we can conclude that photon eigenstate does not contain vacuum average of Higgs fields (see [14-16]) but depends from the hypercharges of Higgs fields (formulae (20) and (21); (see [15-16]). The fixation of hypercharges of fermions fields by the Higgs fields and the dependence of the electric charges quantization conditions from the hypercharges of Higgs fields can be interpreted as influence of Higgs fields on the electric charge quantization. Unlike results of [14] it is shown that in the considered model the classical constraints following from the Yukawa interactions are equivalent to the conditions following from the P – invariance of electromagnetic interaction. This fact may be useful for the explanation of the P – invariance of electromagnetic interaction.